\documentclass[prb,superscriptaddress,twocolumn,floatfix,showpacs]{revtex4-1}
\usepackage{graphicx,color}
\usepackage{amsmath,amssymb,bm}

\newcommand{\tn}{\textnormal}

\definecolor{darkred}{rgb}{0.90,0,0}
\definecolor{darkgreen}{rgb}{0,0.60,.2}
\definecolor{darkblue}{rgb}{0,0,1}
\definecolor{grey}{cmyk}{0,0,0,0.25}
\definecolor{orange}{cmyk}{0,0.6,0.8,0}

\begin{document}
\title{\boldmath The Drude weight of the spin-1/2 $XXZ$  chain:\\ density matrix renormalization group versus exact diagonalization}

\author{C.\ Karrasch}
\affiliation{Department of Physics, University of California, Berkeley, California 95720, USA}
\affiliation{Materials Sciences Division, Lawrence Berkeley National Laboratory, Berkeley, CA 94720, USA}
\author{J. Hauschild}
\affiliation{Department of Physics and Arnold Sommerfeld Center for Theoretical Physics,
Ludwig-Maximilians-Universit\"at M\"unchen, D-80333 M\"unchen, Germany}
\author{S. Langer}
\affiliation{Department of Physics and Arnold Sommerfeld Center for Theoretical Physics,
Ludwig-Maximilians-Universit\"at M\"unchen, D-80333 M\"unchen, Germany}
\affiliation{Department of Physics and Astronomy, University of Pittsburgh, Pittsburgh, Pennsylvania 15260, USA}
\author{F. Heidrich-Meisner}
\affiliation{Department of Physics and Arnold Sommerfeld Center for Theoretical Physics,
Ludwig-Maximilians-Universit\"at M\"unchen, D-80333 M\"unchen, Germany}
\affiliation{Institut f\"ur Theoretische Physik II, Friedrich-Alexander Universit\"at Erlangen-N\"urnberg, Erlangen, Germany}

\begin{abstract}

We revisit the problem of the spin Drude weight $D$ of the integrable spin-$1/2$ $XXZ$  chain using two complementrary approaches, exact diagonalization (ED) and the time-dependent density-matrix renormalization group (tDMRG). We pursue two main goals.  First, we present extensive results for the temperature dependence of $D$. By exploiting time translation invariance within tDMRG, one can extract $D$ for significantly lower temperatures than in previous tDMRG studies. Second, we discuss the numerical quality of the tDMRG data and elaborate on details of the finite-size scaling of the ED results,  comparing calculations carried out in the canonical and grand-canonical ensembles. Furthermore,
we  analyze the behavior of the Drude weight as the point with  SU(2)-symmetric exchange is approached and discuss the relative contribution of the Drude weight 
to the sum rule as a function of temperature.
\end{abstract}

\pacs{71.27.+a, 75.10.Pq, 75.40.Mg, 05.60.Gg}
\maketitle

\section{Introduction}
\label{sec:intro}
Due to the existence of many powerful theoretical methods for one-dimensional systems such as bosonization,\cite{giamarchi}
the density matrix renormalization group,\cite{schollwoeck05,schollwoeck11} or approaches that
exploit integrability,\cite{kluemper-review} many properties of strongly correlated one-dimensional
systems are well studied. A paradigmatic model is the spin-1/2 $XXZ$ chain which maps to interacting 
spinless fermions via the Jordan-Wigner transformation.\cite{jordan28} While  ground-state and
thermodynamic properties (see, e.g., Refs.~\onlinecite{eggert94,kluemper00}) as well as many static and dynamical response functions (see, e.g., Refs.~\onlinecite{klauser11,pereira06}) are very well understood,
a complete theory of transport in this model still poses a formidable challenge to theorists.
A question that has attracted considerable attention for a long time is whether spin transport at finite temperature is 
ballistic or diffusive and whether or  not ballistic transport properties are
related to integrability through the existence of non-trivial local conservation laws.\cite{zotos97,castella95}
Moreover, weakly violated conservation laws can also cause anomalous transport properties with long relaxation rates in one-dimensional systems.\cite{rosch00,jung06}
For the most part, such questions have been studied in the framework of linear response theory (see Refs.~\onlinecite{zotos-review,hm07} and references therein),
yet more recent work has also addressed the non-equilibrium regime (see, e.g., Refs.~\onlinecite{gobert05,prosen09,langer09,benenti09,langer11,jesenko11,znidaric11,znidaric11a,karrasch12a}).
Spin-transport in different experimental realizations of 
quasi one-dimensional spin systems has been probed indirectly by
NMR\cite{takigawa_dynamics_1996,thurber_^17o_2001,pratt_low-temperature_2006}
and $\mu$-SR\cite{master_low_2012}  experiments. Yet, there is no consistent interpretation
of these various experiments
with regard to the question of ballistic or diffusive spin dynamics.
The latest results for
SrCuO$_2$, though, are consistent with  ballistic 
dynamics at elevated temperatures.\cite{master_low_2012}

Here we revisit the problem of finite-temperature spin transport in the spin-1/2 $XXZ$ chain by using
two complementary methods, exact diagonalization and the time-dependent density matrix renormalization group.\cite{vidal04,white04,daley04} 
The Hamiltonian is given by:
\begin{equation}
H = J_H \sum_{i} \lbrack S^x_iS^x_{i+1} + S^y_iS^y_{i+1} + \Delta S^z_iS^z_{i+1}\rbrack,
\label{eq:ham}
\end{equation}
where $S^{\mu}_i$, $\mu=x,y,z$, are the components of a spin-1/2 operator acting on site $i$ of
a chain of length $L$ (boundary conditions will be specified later). $\Delta$ parameterizes the 
exchange anisotropy, $J_H$ sets the energy scale and will be set to unity hereafter ($\hbar=1$). 
The model has a critical gapless phase for $|\Delta| \leq 1$ and  gapped  phases
with antiferromagnetic order and ferromagnetic order for $\Delta>1$ and $\Delta<-1$, respectively.\cite{kolezhuk-review}
We will focus on the gapless phase $|\Delta| \leq 1$ and study spin transport in the linear response regime.

Within linear response, one decomposes the real part of the spin conductivity $\sigma(\omega)$ into 
a singular, zero-frequency part weighted with the so-called Drude weight $D$ and a regular contribution at finite frequencies, $\sigma_\tn{reg}(\omega)$:
\begin{equation}
\tn{Re } \sigma(\omega) = 2\pi D \delta(\omega) + \sigma_\tn{reg}(\omega)\,.
\end{equation}
Note that the present notation for $D$ differs from the one of Ref.~\onlinecite{hm03} by a factor of $2\pi$. We define transport at finite temperatures to be ballistic if $D(T)>0$ and our work focusses on the dependence of $D$ on temperature and exchange anisotropy $\Delta$.
A regular part with a nonzero contribution to the total weight $I_0 =\int d\omega\, \tn{Re } \sigma(\omega)$
is present for $\Delta \not=0$ and $|\Delta|<\infty$. Its precise low-frequency behavior determines whether
this regular channel is diffusive or not (see, e.g., Refs.~\onlinecite{naef98,prelovsek04,sirker09,sirker11,grossjohann10,steinigeweg09,steinigeweg12,herbrych12}). 
More recently, the momentum dependence of current response functions\cite{steinigeweg12} as well as the transverse component of the spin current correlations\cite{steinigeweg11}
have also been investigated. 

For the Drude weight, most studies agree that it vanishes in the massive regime.\cite{zotos96,peres99,hm03,prelovsek04,sirker11,sachdev}
At $\Delta=0$, $D$ is finite for any temperature since the model maps onto non-interacting fermions. 
For $T=0$, $D$ can be obtained exactly via Bethe ansatz.\cite{shastry90} 
Very involved generalizations of the Bethe ansatz to finite temperatures\cite{zotos99,benz05} consistently yield $D>0$ for all $0\leq\Delta<1$; however, the results of 
Refs.~\onlinecite{zotos99,benz05} differ quantitatively regarding the temperature dependence of $D$ and the point of SU(2)-symmetric exchange, $\Delta=1$ (see the discussion in Refs.~\onlinecite{benz05,sirker09,sirker11}). Both approaches have also been argued to violate exact relations at infinite temperatures.\cite{benz05,sirker11}

For nonzero magnetization $S^z=\sum_{i=1}^L S_i^z$, a lower  bound to $D$,
\begin{equation}
0<D_{\mathrm{bound}} \leq D\,,\label{eq:bound}
\end{equation}
can be easily constructed from local conserved quantities known from the Bethe ansatz,\cite{zotos97} 
exploiting the existence of the so-called Mazur inequality. A finite bound requires a non-zero overlap $\langle J Q\rangle$  of the current operator $J$ with at least one conserved quantity $Q$. The existence of such non-zero bounds has been used in Refs.~\onlinecite{zotos97,hm05,sirker09,sirker11} to study the temperature and $\Delta$-dependence
of $D$ at finite magnetizations.
For $S^z=0$, however, the overlap of all local conserved  quantities known from the Bethe ansatz solution with the current operator vanishes for symmetry reasons.\cite{zotos97} Only recently, a {\it non-zero} bound $D_{\mathrm{bound}}>0$ was derived by Prosen for the case of vanishing total magnetization $S^z=0$ and $0<\Delta<1$.\cite{prosen11} This bound exploits the existence of a quasi-local conserved quantity that has a finite overlap with $J$, which can be viewed as the  formal reason as to why $D(T)$ is finite (see also Ref.~\onlinecite{ilievski12} for an extended discussion). 
It remains unclear whether or not this conserved
quantity is related to integrability and whether the bound presented in Ref.~\onlinecite{prosen11} is exhaustive [equality in Eq.~\eqref{eq:bound}] at all $0<\Delta<1$. 
Despite this great progress in understanding why $D(T)$ is nonzero in this model, a result that previously was mostly based on numerical simulations,\cite{zotos96,narozhny98,hm03,hm07,heidarian07,herbrych11,karrasch12} there are still relevant open questions. First, at $\Delta=1$, there is no agreement as to whether or not $D$ is finite. Second, the temperature dependence of $D(T)$ for $|\Delta|<1$, albeit studied in previous work,\cite{zotos99,benz05,fujimoto03,alvarez02,hm03,herbrych11,heidarian07,karrasch12} 
is not fully understood yet and is closely related to the question of which fraction of the total weight $I_0$ is in the Drude weight compared to finite-frequency
contributions.\cite{sirker09,sirker11,herbrych12} Finally, when evaluating expressions for $D$ with exact diagonalization, one can either elect to work
in a grand-canonical picture (i.e., averaging over all subspaces with a given total $S^z$) or in the canonical ensemble (i.e., only one sector with a fixed 
$S^z$ is taken into account). While one would, on general grounds,  expect that in the thermodynamic limit both approaches should yield the same result,
a vastly  different behavior can emerge on finite systems.\cite{herbrych11,hm03} This calls for a more detailed investigation.  

The first method that we employ  is the time-dependent density matrix renormalization group for finite temperatures. Various approaches for using tDMRG at finite temperatures
have been suggested and pursued in the literature.\cite{sirker05,barthel09,white09,feiguin05,verstraete04,karrasch12,barthel13a,barthel13b}
The main restriction of this method, which can otherwise provide highly accurate results, is the maximally accessible time scale
due to the entanglement encoded in the time-evolved wave-function (see Ref.~\onlinecite{schollwoeck11} for a discussion).
Here, we use the purification method originally proposed in Ref.~\onlinecite{verstraete04} complemented by a {\it disentangler} that allows one to go
to longer times than previously possible.\cite{karrasch12} In a nutshell, the purification approach relies on embedding the actual physical system for which one seeks to know
finite-temperature properties into some environment (which in our case is simply a copy of the physical system). One can then show that a pure-state time-evolution of the full system, which is amenable to tDMRG, yields the 
thermal statistical operator for the system after tracing out the environment's degrees of freedom.
An advantage of this tDMRG approach is that the results for the accessible time-scales are virtually in the 
thermodynamic limit. In Ref.~\onlinecite{karrasch12}, this method was used to extract the Drude weight
of the spin-1/2 $XXZ$ chain from the time dependence of current-current autocorrelation functions. The results of Ref.~\onlinecite{karrasch12}
confirm the prevailing picture that $D(T)>0$ for $|\Delta|<1$ while for $\Delta =1$, the accessible time scales are too short to arrive
at definite conclusions. Here, we provide a more detailed error analysis of this method and we present extensive results for the 
temperature-dependence of $D(T)$ for $|\Delta| <1$, in particular including  $\Delta <0$. It was recently suggested\cite{barthel13a,barthel13b} to rewrite a correlation function as $\langle A(t)A\rangle=\langle A(t/2)A(-t/2)\rangle$, which allows to reach times twice as large without much additional programming effort. By exploiting this trick we can extract the Drude weight via tDMRG for significantly lower temperatures than the ones studied in Ref.~\onlinecite{karrasch12}.

Our second method is exact diagonalization which allows to compute transport coefficients for typically $L\lesssim 20$ sites.\cite{hm03,hm07,herbrych11}
The expression for the Drude weight is:\cite{shastry90,castella95}
\begin{equation}
D = \frac{1}{2L}\left\lbrack \langle  - \hat T\rangle -\frac{2}{Z_T} \sum_{m, n\atop E_m\not=E_n}
 e^{-\beta E_n}\frac{|\langle m |
J|n\rangle|^2}{E_m-E_n} \right\rbrack\,,\label{eq:drude}
\end{equation}
where $Z_T=\sum_n \mbox{exp}(-\beta E_n)$ is the partition function,  $| n\rangle$ and $E_n$ are the eigenstates and eigenvalues of $H$, $J$ is the spin-current operator,
and $\hat T= \sum_i (S_i^x S_{i+1}^x +S_i^y S_{i+1}^y)$ is the kinetic energy.
We present a detailed comparison of a grand-canonical versus a canonical evaluation of Eq.~\eqref{eq:drude}.
A canonical evaluation of Eq.~\eqref{eq:drude} was  systematically used in a recent work\cite{herbrych11} by Herbrych {\it et al.} to
compute the dependence of $D$ on temperature  and exchange anisotropy.
We argue that over a wide range of parameters ($\Delta$ and temperature $T$) the grand-canonical and canonical extrapolations
yield quantitatively similar results. 
Moreover, we observe that the {\it grand-canonical} data often exhibits weaker finite-size dependencies than the canonical ones, agrees better with tDMRG data, and it seems to depend on system size in 
a  more systematic way throughout the parameter range of interest, i.e.,
$|\Delta|\leq 1$. Furthermore, it has long been realized that results for $D$ from systems with an odd number of sites versus an even number of sites
exhibit a different finite-size dependence,\cite{hm03,jung07,herbrych11} which appears to be related to non-trivial finite-size effects 
of the regular part $\sigma_\textnormal{reg}(\omega)$ at small frequencies.\cite{naef98,rigol08,herbrych11} The interplay of the Drude weight and finite-frequency contributions was
systematically studied by Herbrych {\it et al.}\cite{herbrych11} who show that these effects have
to be taken into account into the finite-scaling at incommensurate $\Delta$
(i.e., those values of $\Delta$ that cannot be written as $\Delta=\cos(\pi/\nu)$, $\nu$ integer).  These finite-size dependencies can be understood by considering  that the Drude weight is related to the curvature of many-body levels upon changing boundary conditions via a flux.\cite{castella95,zotos97,herbrych11} To avoid this complication, we focus on odd system sizes and commensurate values of $\Delta$.\cite{naef98} Under these three premises, grand-canonical evaluation of Eq.~\eqref{eq:drude}, consideration of odd $L$, commensurate values of $\Delta$,
we analyze the temperature dependence of $D(T)$ throughout the gapless phase and we present results for $D(T)$ extrapolated to the thermodynamic limit.
Unfortunately, there is no theory for the finite-size dependence of $D(T)$ which is the main drawback of the exact diagonalization approach.
However, for those parameters for which reliable results from ED and tDMRG are available, we report an excellent quantitative agreement between
these two methods.
To elucidate the behavior in the vicinity of $\Delta=1$, we follow $D=\cos(\pi/\nu)$ as a function of $\nu$ and observe that
$D(\nu)$ seems to approach finite values as $\nu\to\infty$ (corresponding to $\Delta=1$).

\begin{figure}[t]
\includegraphics[width=0.95\linewidth,clip]{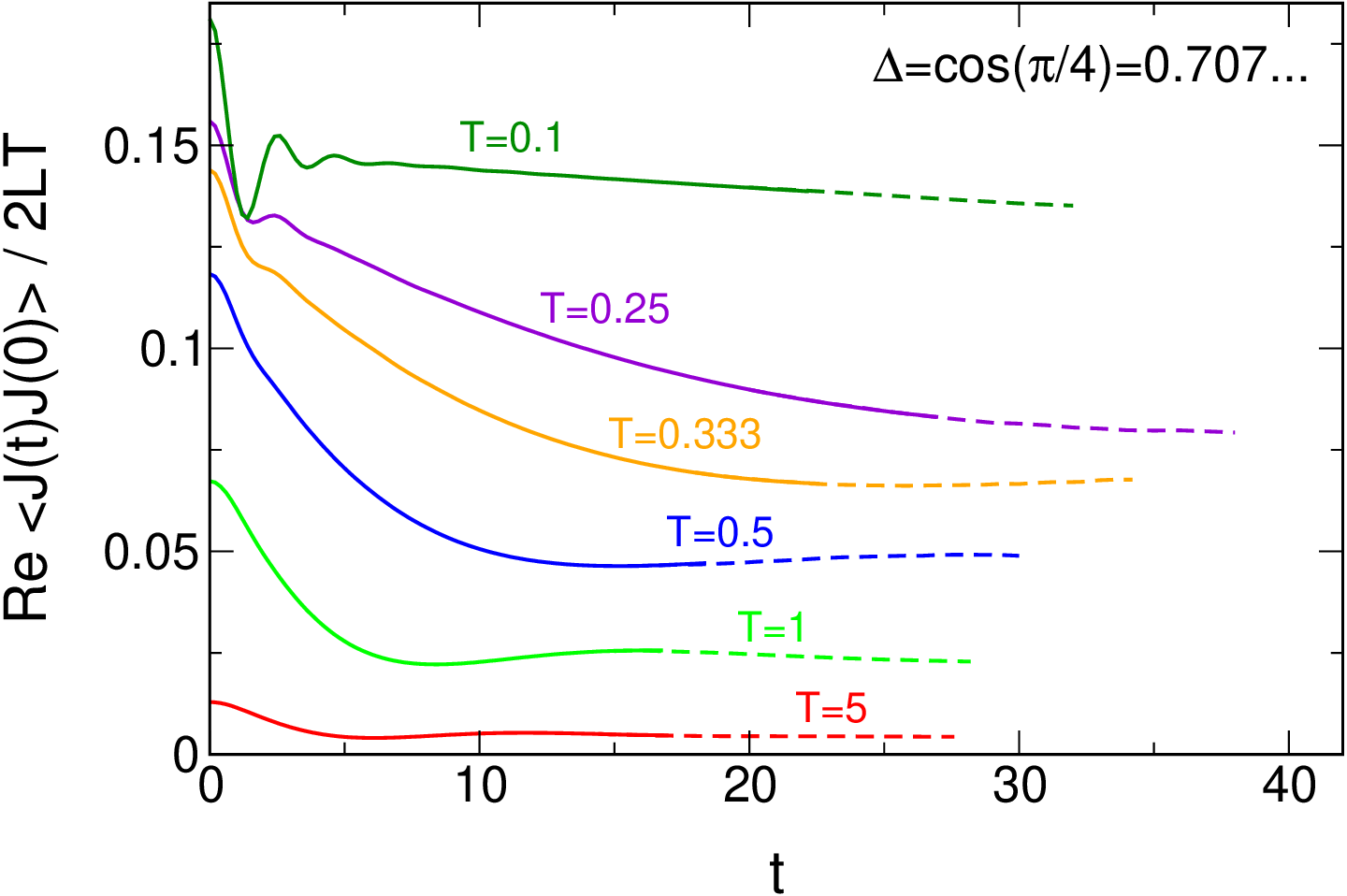}
\caption{(Color online) DMRG data for the real-time current-current correlation function of an $XXZ$  chain with $\Delta=0.707$ at intermediate temperatures $T \leq 5$. Its asymptotic value at long times  determines the Drude weight through Eq.~(\ref{eq:drude1}). The calculation was carried out for  $L=200$ and using a discarded weight $\epsilon=10^{-5}$ during the real time evolution. Dashed lines were obtained by exploiting Eq.~(\ref{eq:curr2}).}
\label{fig:trick}
\end{figure}

To summarize, the main results of our work are an extensive analysis of $D=D(T)$ using both tDMRG and ED which are in very good agreement, a discussion
of the numerical quality of the tDMRG data, and a detailed comparison of the finite-size dependencies of ED data. 
Exploiting time translation invariance allows to access lower temperatures than in previous tDMRG studies.\cite{karrasch12}

The structure of this exposition is as follows. In Sec.~\ref{sec:coeffs}, we summarize the expressions for the transport coefficients that are studied 
in this work. In Sec.~\ref{sec:tdmrg}, we discuss details of the tDMRG approach while Sec.~\ref{sec:ed} focusses on aspects specific to ED.
We present our main results and a discussion thereof in Sec.~\ref{sec:results} and conclude with a brief summary, contained in Sec.~\ref{sec:sum}.    

\section{Definitions of transport coefficients}
\label{sec:coeffs}

\begin{figure}[t]
\includegraphics[width=0.95\linewidth]{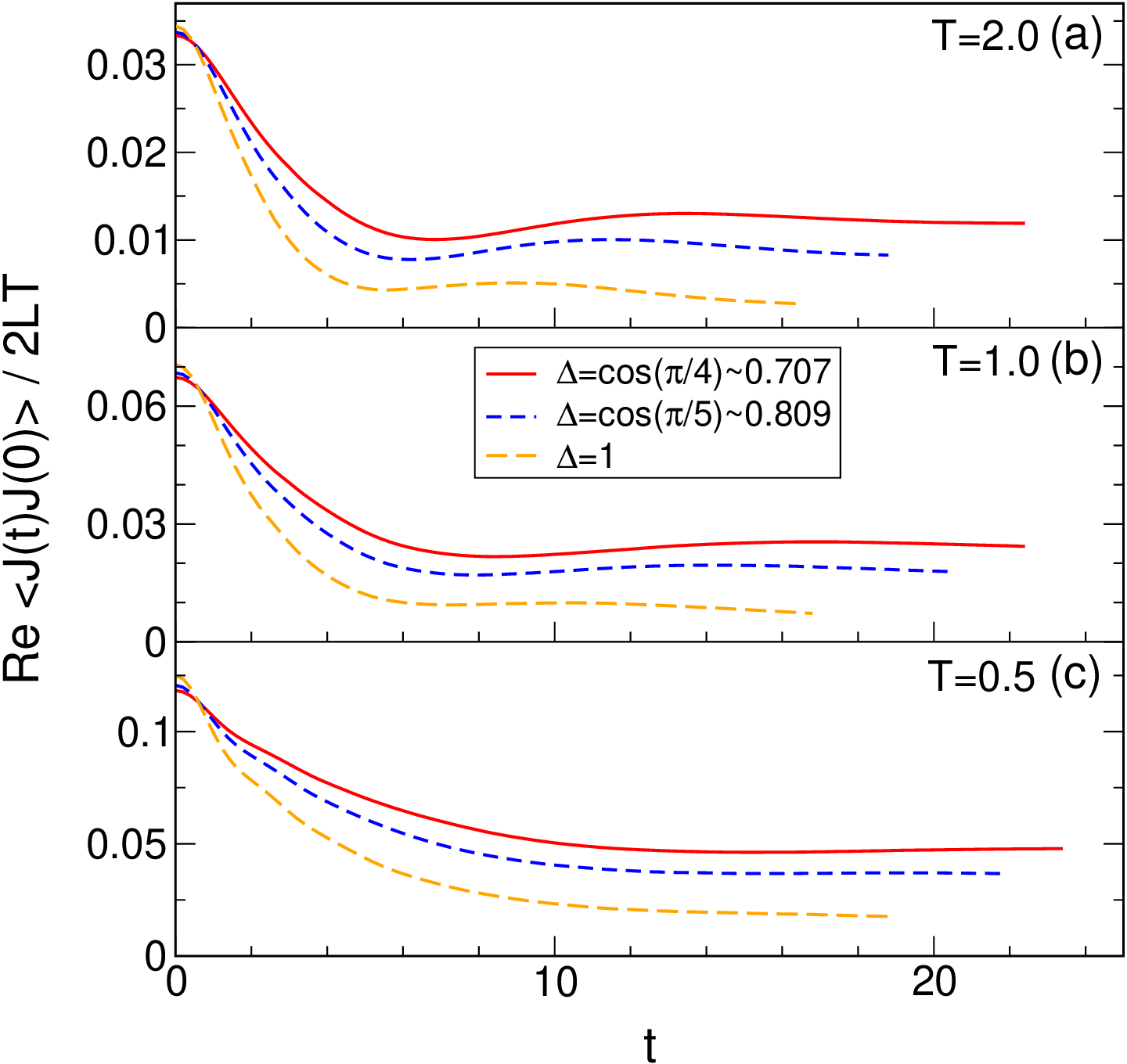}
\caption{(Color online) Current-current correlation function for various anisotropies $\Delta>0$. The calculation was carried out using $L=100$, $\epsilon=10^{-4}$ without exploiting Eq.~(\ref{eq:curr2}). }
\label{fig:corr1}
\end{figure}

Our aim is to compute the Drude weight of the integrable $XXZ$  chain defined in Eq.~\eqref{eq:ham}. 
The Kubo formula for the spin conductivity reads
\begin{equation}\begin{split}
\sigma(\omega) & = \frac{1}{\omega L}\int_0^\infty e^{i\omega t} \langle [J(t),J(0)]\rangle dt \\
& = \frac{1-e^{-\omega/T}}{\omega L} \int_0^\infty e^{i\omega t} \langle J(t)J(0)\rangle dt~,
\end{split}\end{equation}
where the finite-temperature real-time current-current correlation function is given by
\begin{equation}\label{eq:corr}
\langle J(t) J(0) \rangle = \tn{Tr}\left( \rho_T\, e^{iHt} J e^{-iHt} J\right) ~,~~\rho_T = \frac{e^{-H/T}}{Z_T}~.
\end{equation}
 The current operator $J=\sum_i j_i$ is defined via a continuity equation
\begin{equation}
\partial_t S_i^z = j_{i}-j_{i+1} ~~\Rightarrow~~  j_i = i S^+_iS^-_{i+1}/2 + \tn{H.c.}~~.
\end{equation}

By inserting a full set of eigenstates $| n\rangle$ and eigenvalues $E_n$ in Eq.~\eqref{eq:corr},  the Drude weight can  be
cast  into Eq.~\eqref{eq:drude}.\cite{shastry90,castella95}
This expression can be evaluated using exact diagonalization.

The Drude weight is further related to the long-time asymptote of of the current-current correlation function $\langle J(t) J(0)\rangle$ (see Ref.~\onlinecite{zotos97}):
\begin{equation}\label{eq:drude1}
D = \lim_{t\to\infty}\lim_{L\to\infty} \frac {\textnormal{Re } \langle J(t)J(0)\rangle}{2LT}~.
\end{equation}
Using time-dependent DMRG, the current-current correlation function can be computed, thus yielding access to the
Drude weight.

\begin{figure}[t]
\includegraphics[width=0.95\linewidth,clip]{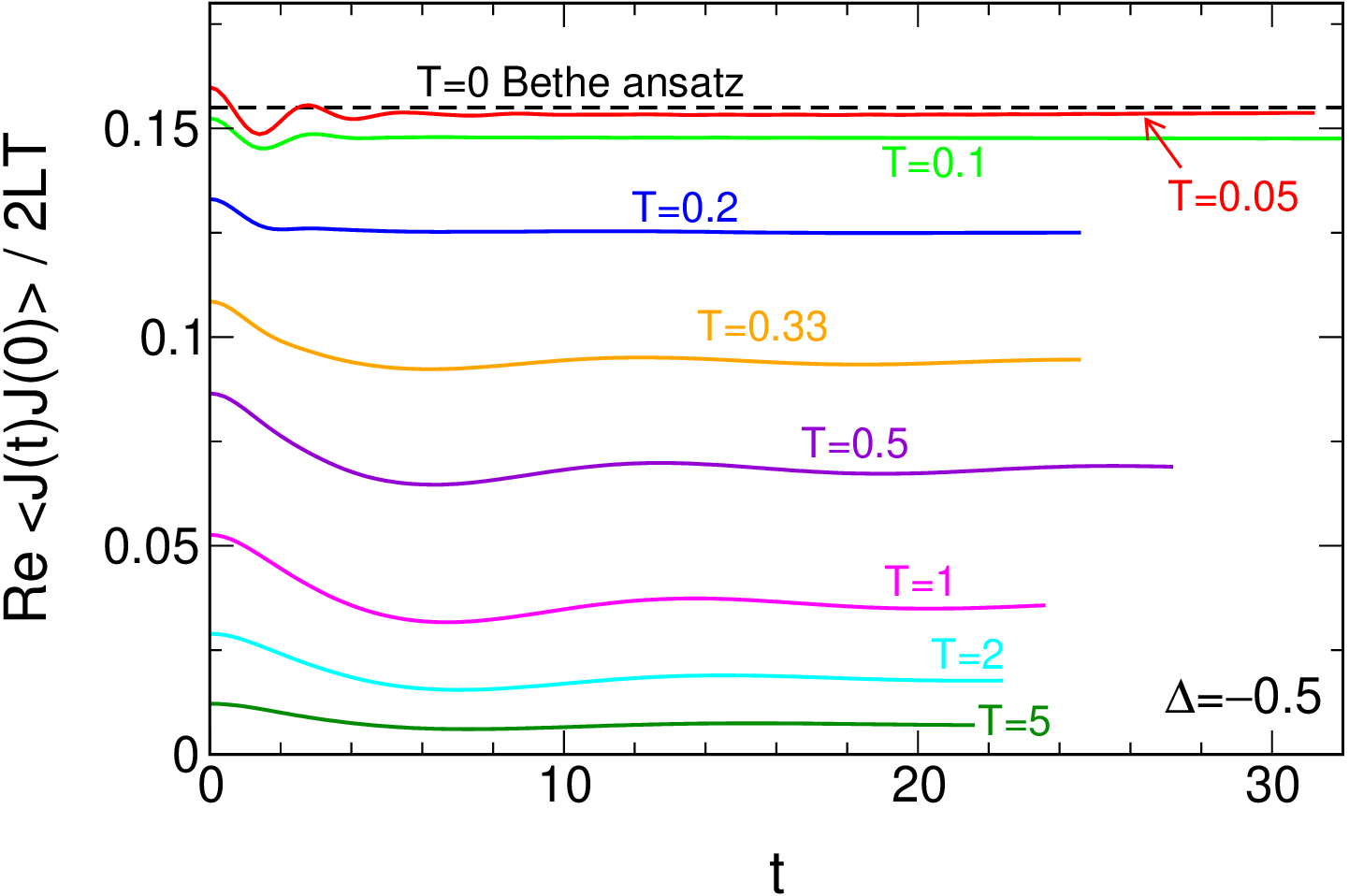}
\caption{(Color online) The same as in Fig.~\ref{fig:corr1} but for $\Delta<0$ and over the full range of  temperatures $0.05 \leq T\leq 5$. For $T\to0$ one approaches the $T=0$ Bethe ansatz result.\cite{shastry90} The calculations were carried out using $L=200$ as well as $\epsilon=10^{-5}$ for $T=\{5,2,1,0.5\}$, $\epsilon=10^{-6}$ for $T=\{0.33,0.2\}$, and $\epsilon=10^{-7}$ for $T=\{0.1,0.05\}$.}
\label{fig:corr2}
\end{figure}

\section{Density matrix renormalization group simulations}
\label{sec:tdmrg}

\subsection{Method}

In order to evaluate Eq.~(\ref{eq:corr}) by a DMRG algorithm\cite{white92,white93,schollwoeck11} that evolves a wave-function (as opposed to the time-evolution of a matrix product operator), one first needs to purify the thermal density matrix $\rho_T$ by introducing an auxiliary Hilbert space $Q$ such that  $\rho_T = \tn{Tr\,}_{Q} |\Psi_T\rangle\langle\Psi_T|$. This is analytically possible only at $T=\infty$ where $\rho_T$ factorizes. However, $|\Psi_T\rangle$ can be obtained from $|\Psi_\infty\rangle$ by applying an imaginary time evolution, $|\Psi_T\rangle=e^{-H/(2T)}|\Psi_\infty\rangle$.\cite{verstraete04,feiguin05,barthel09} The current-current correlation function of Eq.~(\ref{eq:corr}) can then be \textit{exactly} recast as
\begin{equation}\label{eq:curr1}
\langle J(t)J(0)\rangle = \frac{\langle \Psi_\infty| e^{-H/2T}e^{iHt} J e^{-iHt} J e^{-H/2T}|\Psi_\infty\rangle}{\langle\Psi_\infty| e^{-H/T}|\Psi_\infty\rangle}~,
\end{equation}
and this object is directly accessible in the time-dependent DMRG framework\cite{white04,daley04,vidal04} outlined in Ref.~\onlinecite{karrasch12}. In DMRG simulations, we use
open boundary conditions.

It is convenient to first express $|\Psi_\infty\rangle$ in terms of a matrix product state
\begin{equation}\label{eq:mps}
|\Psi_\infty\rangle = \sum_{\sigma_n} A^{\sigma_0}A^{\sigma_1}\cdots A^{\sigma_{2L-1}}|\sigma_0\sigma_1\ldots\sigma_{2L-1}\rangle~,
\end{equation}
where
\begin{align}
A^{\uparrow_\tn{even}}& = (1~~0) &
A^{\downarrow_\tn{even}}& = (0~~-1) \nonumber \\
A^{\uparrow_\tn{odd}}& = (0~~1/\sqrt{2})^T &
A^{\downarrow_\tn{odd}}& = (1/\sqrt{2}~~0)^T ~.
\end{align}
Physical (auxiliary) sites are denoted by even (odd) indices. After factorizing the evolution operators $\exp(-\lambda H)$ using a fourth-order Trotter decomposition, they can be successively applied to Eq.~(\ref{eq:mps}). At each time-step $\Delta\lambda$, two singular value decompositions are carried out to update three consecutive matrices. The matrix dimension $\chi$ is dynamically increased such that at each time step the sum of all squared discarded singular values is kept below a threshold value $\epsilon$. As mentioned above, we do not only time-evolve the physical but also the auxiliary sites (using the physical Hamiltonian but with reversed time). As outlined in Ref.~\onlinecite{karrasch12}, this is an exact modification to the DMRG algorithm (auxiliary sites are traced over, and one can thus apply an arbitrary unitary transformation to them) which leads to a significantly slower build-up of entanglement; hence, longer time scales can be reached. The tDMRG approximation to $\langle J(t)J(0)\rangle$ becomes exact in the limit of $\epsilon\to0$ and $\Delta\lambda\to0$. We ensure that both are chosen small enough and that the system size $L$ is large enough to obtain results which are converged w.r.t.~those parameters (this is discussed below). 

The main cost of time-dependent DMRG simulations, both at finite and zero temperature, depends on the entanglement growth as a function of time.\cite{schollwoeck11}
For the methods used here, this problem was discussed  in Refs.~\onlinecite{barthel13b} and \onlinecite{karrasch13} which we refer the reader to for details and
examples. 

\begin{figure}[t]
\includegraphics[width=0.95\linewidth,clip]{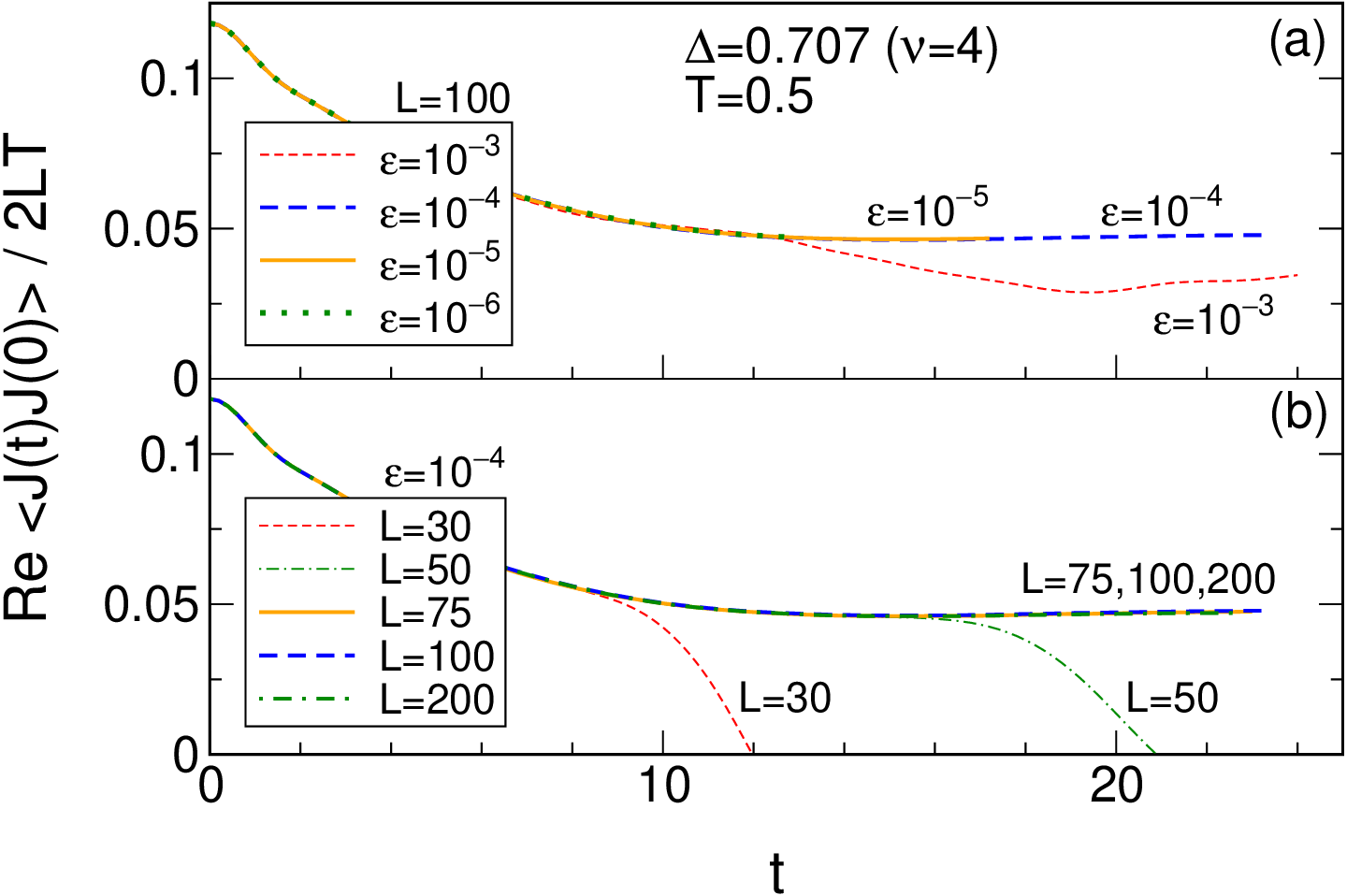}
\caption{(Color online) DMRG data for the current-current correlation function obtained using (a) different discarded weights $\epsilon$ during the real-time evolution and (b) different system sizes $L$ ($\nu=4$, $T=0.5$). }
\label{fig:params}
\end{figure}

In Refs.~\onlinecite{barthel13a,barthel13b} it was recently suggested to exploit time translation invariance to rewrite any correlation function as
\begin{equation}\label{eq:curr2}
\langle A(t)A(0)\rangle = \langle A(t/2)A(-t/2)\rangle~.
\end{equation}
This trick allows to access times \textit{twice as large} without much additional effort. Within the framework of purification, one needs to compute\cite{karrasch13}
\begin{equation}
e^{-iHt/2} A e^{iHt/2} e^{-H/2T}|\Psi_\infty\rangle~.
\end{equation}
It is most straightforward to calculate $\langle J(t/2)J(-t/2)\rangle$ by evaluating every term $j_i$ that contributes to $J=\sum_ij_i$ separately. Thus, the effort to obtain the current-current correlation function via $\langle J(t/2)J(-t/2)\rangle$ is larger (but not exponentially larger) than it is for $\langle J(t)J(0)\rangle$ where it is sufficient to carry out a single time evolution $e^{-iHt}j_{L/2}e^{-H/2T}|\Psi_\infty\rangle$.\cite{sirker11} We therefore mostly stick to Eq.~(\ref{eq:curr1}) and only exemplary (i.e., for a single anisotropy $\Delta=0.707$ and temperatures $T=0.25\ldots5$) exploit Eq.~(\ref{eq:curr2}).

\begin{figure}[t]
\includegraphics[width=0.95\linewidth]{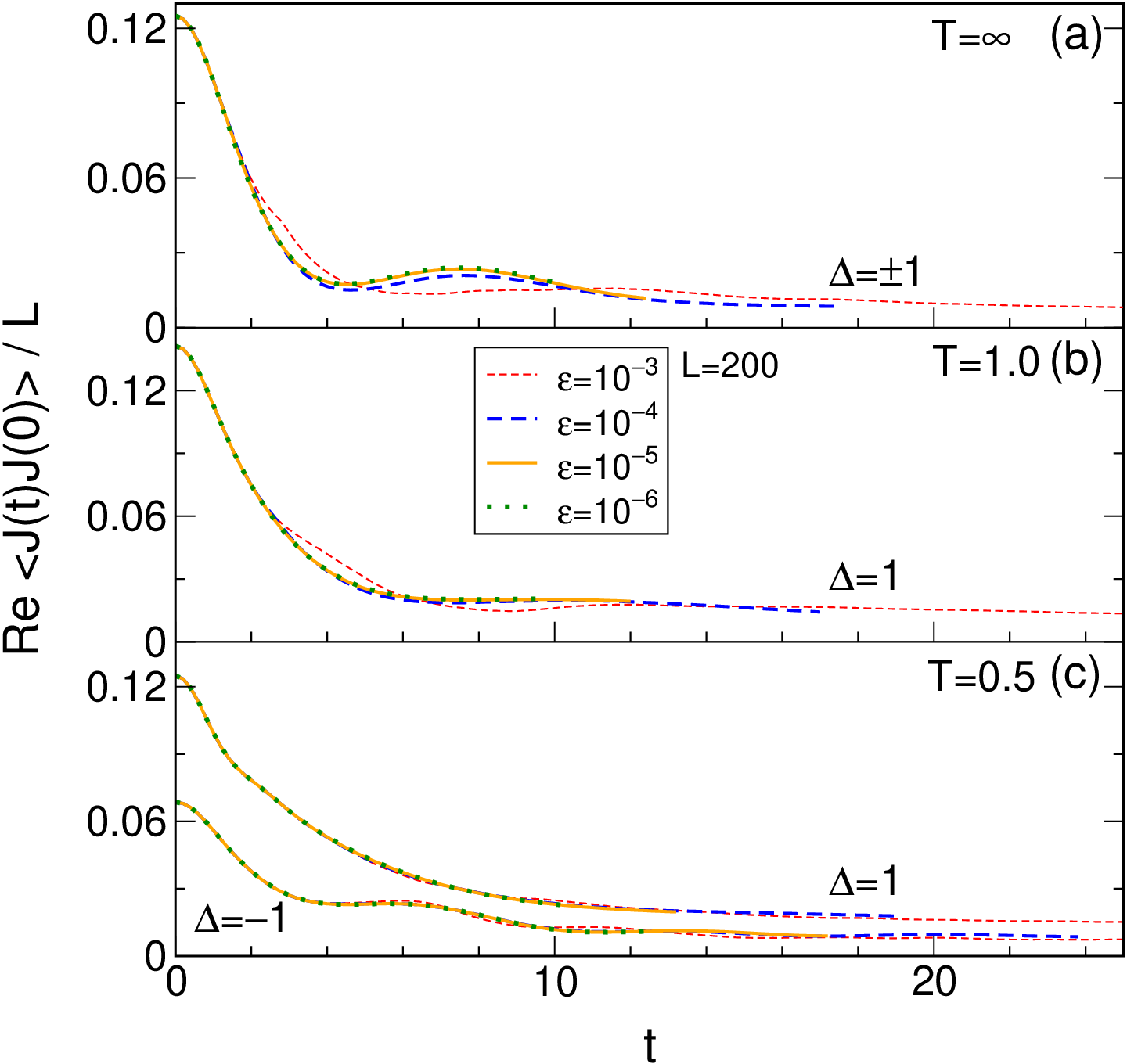}
\caption{(Color online) Error analysis  as a function of various discarded weights for $\Delta=\pm1$, $L=200$ and (a) $T=\infty$, (b) $T=1$ and (c) $T=0.5$. Note that for $T=\infty$ curves at $\pm\Delta$ coincide.\cite{benz05}}
\label{fig:d1m}
\end{figure}

\subsection{Real-time data}

tDMRG data for the current-current correlation function are shown in Figs.~\ref{fig:trick}-\ref{fig:d1m}. The calculation is stopped once the matrix dimension has reached values of $\chi\approx 1000-2000$. This generically allows to access time scales $t\approx20-40$ which are much larger than the intrinsic scale $t=1/J_H=1$. For antiferromagnetic anisotropies $0<\Delta<1$ and intermediate to large temperatures $T\gtrsim0.25$, $\langle J(t)J(0)\rangle$ has saturated for $t\approx40$ (see Fig.~\ref{fig:trick}); thus, the Drude weight can be obtained via Eq.~(\ref{eq:drude1}). At smaller $T$, a large diffusive contribution with a small decay rate prevents us from reaching the asymptotic regime.\cite{sirker09,sirker11} Data for the isotropic chain $\Delta=1$ are shown in Figs.~\ref{fig:corr1} and \ref{fig:d1m}; these will be discussed in Sec.~\ref{sec:gapless}.

In the ferromagnetic case $-1<\Delta < 0$ (Fig.~\ref{fig:corr2}), the current-current correlation function seems to saturate fast for any $T$. Field theory suggests that a diffusive contribution (of unknown total magnitude) with an exponentially small decay rate exists for $-1<\Delta<0$.\cite{sirker09,sirker11} There are thus two possibilities to interpret our data at small $T$: either $\langle J(t)J(0)\rangle$ saturates slowly on a large scale which does not manifest itself for times $t\lesssim20$; or the magnitude of the diffusive contribution is small. The facts  that first, we successively approach the exact Bethe ansatz result for $T\to0$ (see Fig.~\ref{fig:corr2}) and second, the observation of large ballistic contribution relative to the total weight of $\sigma(\omega)$ (compare Sec.~\ref{sec:weight}) support the latter, and in the following we will rely on  this interpretation and discuss the available
numerical data.

As mentioned above, we have to ensure that the discarded weight and the Trotter step size are chosen small enough and that the system size $L$ is chosen large enough to obtain results which are converged w.r.t.~those parameters. Typical values are $L=200$, $\Delta\lambda=0.2$, and $\epsilon=10^{-9}$ as well as $\epsilon=10^{-4}\ldots10^{-7}$ during the imaginary and real-time evolutions, respectively. The dependencies on $L$ and $\epsilon$ are illustrated in Figs.~\ref{fig:params} and \ref{fig:d1m}. The finite system size manifests itself on time scales $t\gtrsim L/v$ with $v$ being a proportionality factor which has the dimension of a velocity. This is the so-called light-cone effect in nonrelativistic systems: correlation functions $\langle A(t)B(0)\rangle$ of local operators $A$ and $B$ fall off exponentially for $x>vt$ with $x$ denoting the spatial distance between the regions on which $A$ and $B$ act.\cite{liebrobinson} Note that in our case for times $t\lesssim20$ and temperatures $T\gtrsim0.5$, the data for all $L>75$ coincides (see Fig.~\ref{fig:params}). Moreover, if $L$ is chosen too small or if $\epsilon$ is chosen too large, one generically \textit{underestimates} the Drude weight.

\section{Exact Diagonalization}
\label{sec:ed}
Our ED analysis of the Drude weight is based on an evaluation of Eq.~\eqref{eq:drude}. We use periodic boundary 
conditions and we compute transport coefficients for  systems as large as $L\leq 19$. 
We will make a distinction between a {\it grand-canonical} versus a {\it canonical} evaluation of Eq.~\eqref{eq:drude},
which includes the evaluation of the kinetic energy from 
\begin{equation}
\langle \hat T\rangle  = \frac{1}{Z} \sum_{n} e^{-\beta E_n} \langle n | \hat T | n\rangle\,. \label{eq:T}
\end{equation}
To be specific about grand- versus canonical interpretation, the sums over eigenstates in Eqs.~\eqref{eq:drude} and \eqref{eq:T} have to
be read as:
$$
\sum_{n} \to \sum_{S^z} \sum_{E_n(S^z)} 
$$
in the former case, whereas in the latter case, we fix $S^z$ to a desired value and sum over all eigenstates $E_n(S^z)$ in that 
subspace only. For odd $L$, the smallest $S^z$ is $S^z=1/2$, which is non-zero (but the magnetization per sites goes to zero as $1/2L$).\cite{herbrych11}

\begin{figure}[t]
\includegraphics[width=0.95\linewidth,clip]{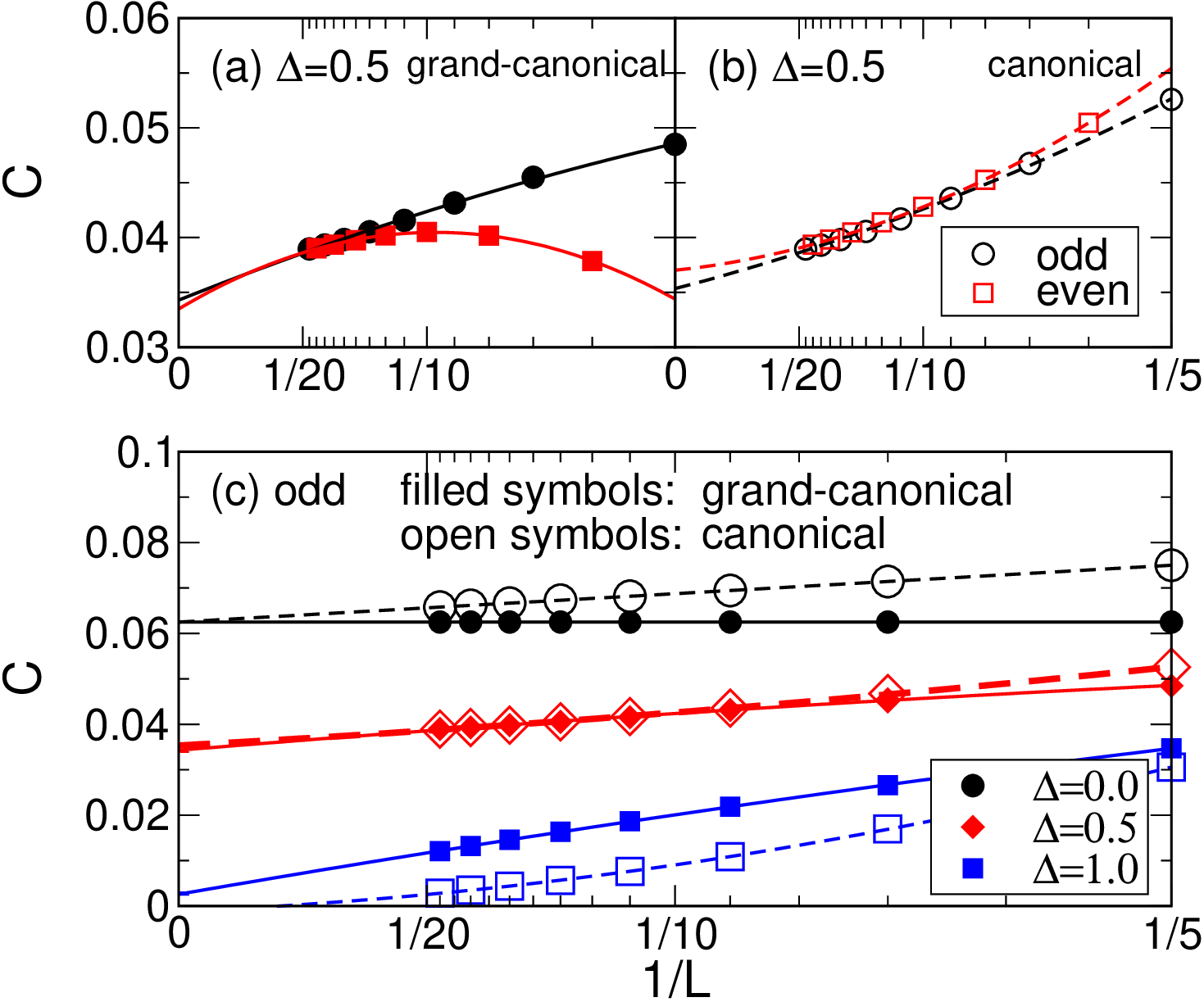}
\caption{
(Color online) ED data for the leading coefficient $C(L)$ of $D(T)\approx C/T$ in the high-temperature limit (see Eq.~\eqref{eq:c}).
(a) Grand-canonical and (b) canonical  data at $\Delta=0.5$, comparing data from odd $L$ to even $L$. The lines are an extrapolation in system size using
$C(L) = C + A/L + B/L^2$ using data-points with $5 \leq L \leq 19$. (c) Comparison between grand-canonical and canonical data for $C$
for odd $L$ and $\Delta=0,0.5,1$. The ED results are extrapolated to $L=\infty$ using Eq.~\eqref{eq:extra_c}.}
\label{fig:ED_extN_C}
\end{figure}

Let us first compare data obtained from chains with an {\it even} versus an {\it odd} number of sites in the example of 
$\Delta=0.5$ and in the high-temperature limit. 
We compute the leading coefficient of $D(T)=C/T+\mathcal{O}(T^{-2})$ in a high-temperature expansion in $1/T$ from\cite{hm03} 
\begin{equation}
C=\lim_{T\to \infty} \lbrack T D(T)\rbrack\,. 
\label{eq:c}
\end{equation}
The results are shown in Fig.~\ref{fig:ED_extN_C}(a) for the grand-canonical case and in Fig.~\ref{fig:ED_extN_C}(b) for the canonical one.
In the latter case, odd system sizes correspond to working with $S^z=1/2$ whereas for even system sizes, we work in the $S^z=1$ subspace (note that $S^z/L\to0$ as $L\to\infty$).\cite{herbrych11}
The figure further contains fits to the data using
\begin{equation}
C(L)= C +A/L+B /L^2\,. \label{eq:extra_c}
\end{equation}
The results for $C$ are all consistent with each other, with a somewhat larger deviation for the canonical data with $S^z=1$.
However, the coefficient $B$ is {\it larger} for chains with an even number of sites, which is particularly evident in the grand-canonical case [see Fig.~\ref{fig:ED_extN_C}(a)], 
indicating that the data for odd system sizes
approaches the large-$L$ limit faster at $T\to \infty$. This behavior extends down to temperatures of $T\sim 0.5$.  In the following, we therefore 
restrict our analysis to odd $L=5,7, \dots, 19$. 

\begin{figure}[t]
\includegraphics[width=0.95\linewidth,clip]{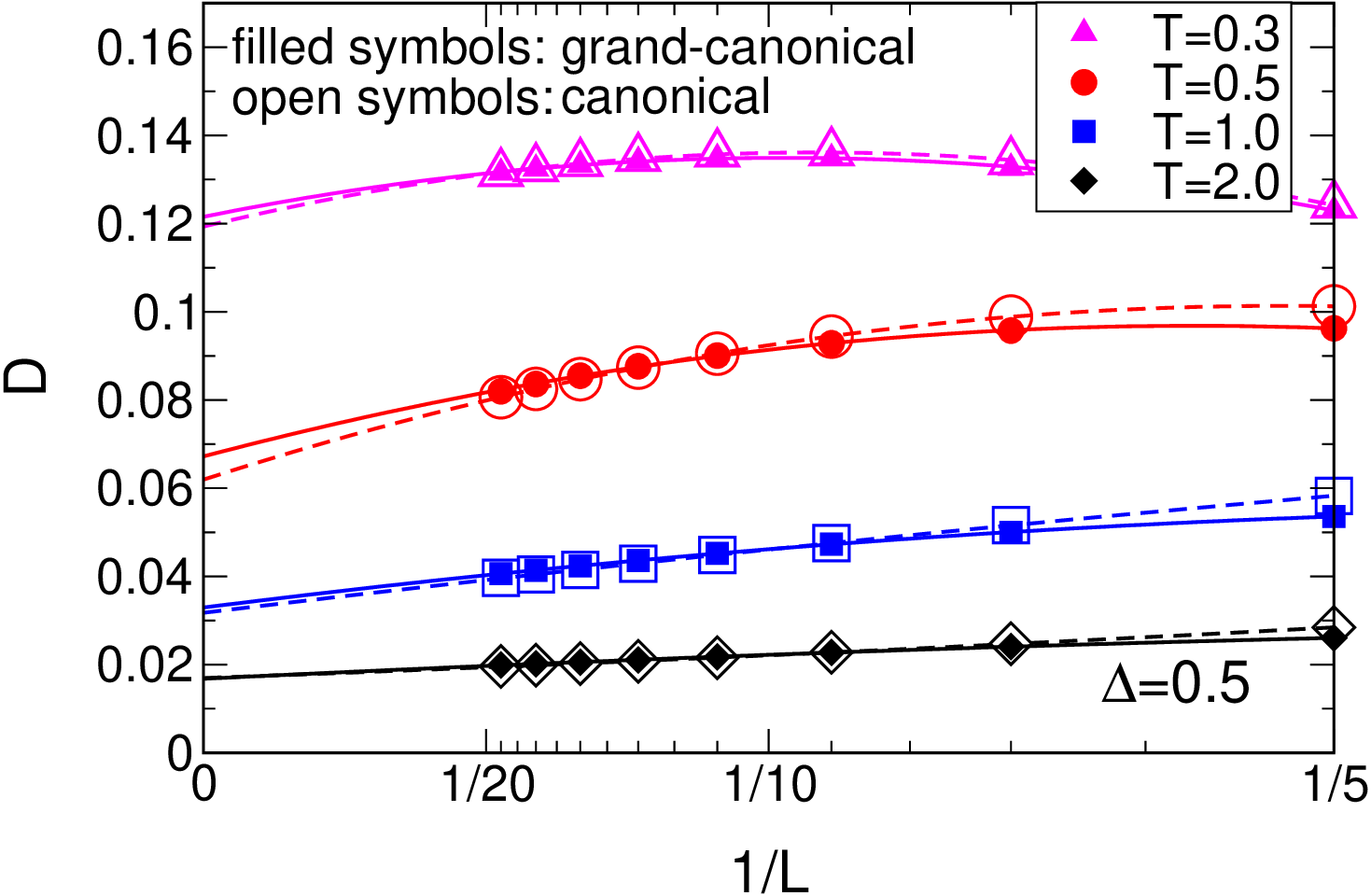}
\caption{
(Color online) ED data for the Drude weight $D$ for $\Delta=0.5$.  The filled (open) symbols are the results for the grand-canonical (canonical) calculation of $D$.
The solid (dashed) lines are an extrapolation of $D$ in system-size using  a second-order polynomial in $1/L$, Eq.~\eqref{eq:d_l}, for grand-canonical (canonical) data.  }
\label{fig:ED_extN_D}
\end{figure}

Next we compare grand-canonical data to canonical data  for $C$ for $\Delta=0,0.5,1$ which is shown in Fig.~\ref{fig:ED_extN_C}(c), extrapolated
using Eq.~\eqref{eq:extra_c}.
We observe that in the high-temperature limit,  $B\approx 0$.\cite{hm03}
Generally, one would expect both the grand-canonical and the canonical data sets  to result in the exact same number as $L\to \infty$. This is consistent with our data
for $\Delta=0$. Clearly, though, the grand-canonical data exhibits {\it no} finite-size dependence at all for $\Delta=0$ in contrast
to the canonical data. Therefore, we conclude that data obtained from a canonical evaluation of Eq.~\eqref{eq:drude} approaches the large 
$L$ limit {\it slower} than grand-canonical data.
At $\Delta=0.5$, the grand-canonical data and the canonical data  extrapolate to the same value. At $\Delta=1$, the grand-canonical 
data still follows the simple function given in Eq.~\eqref{eq:extra_c} quite well and extrapolates to a finite value of $C(L=\infty)$ (this reproduces the results
of Refs.~\onlinecite{hm03,hm07}), whereas the canonical 
data do not follow any low-order polynomial in $1/L$.  
The authors of Ref.~\onlinecite{herbrych11} argue that the canonical data for $\Delta=1$ can only be interpreted in terms of $C(L=\infty)=0$.
Our point of view is different: since there is no theory for the finite-size dependence of $D$, one can only use the canonical
data at $\Delta=1$  extracted from the largest $L$ as an {\it upper bound} to the true $D$, assuming that the finite-size dependence
is monotonous. To summarize, while the grand-canonical data can be well fitted to a simple polynomial in $1/L$ of low order 
for all $|\Delta| \leq 1$, it is at present unclear how to extrapolate the canonical data close to $\Delta=1$ to $L\to \infty$.

\begin{figure}[t]
\includegraphics[width=0.95\linewidth,clip]{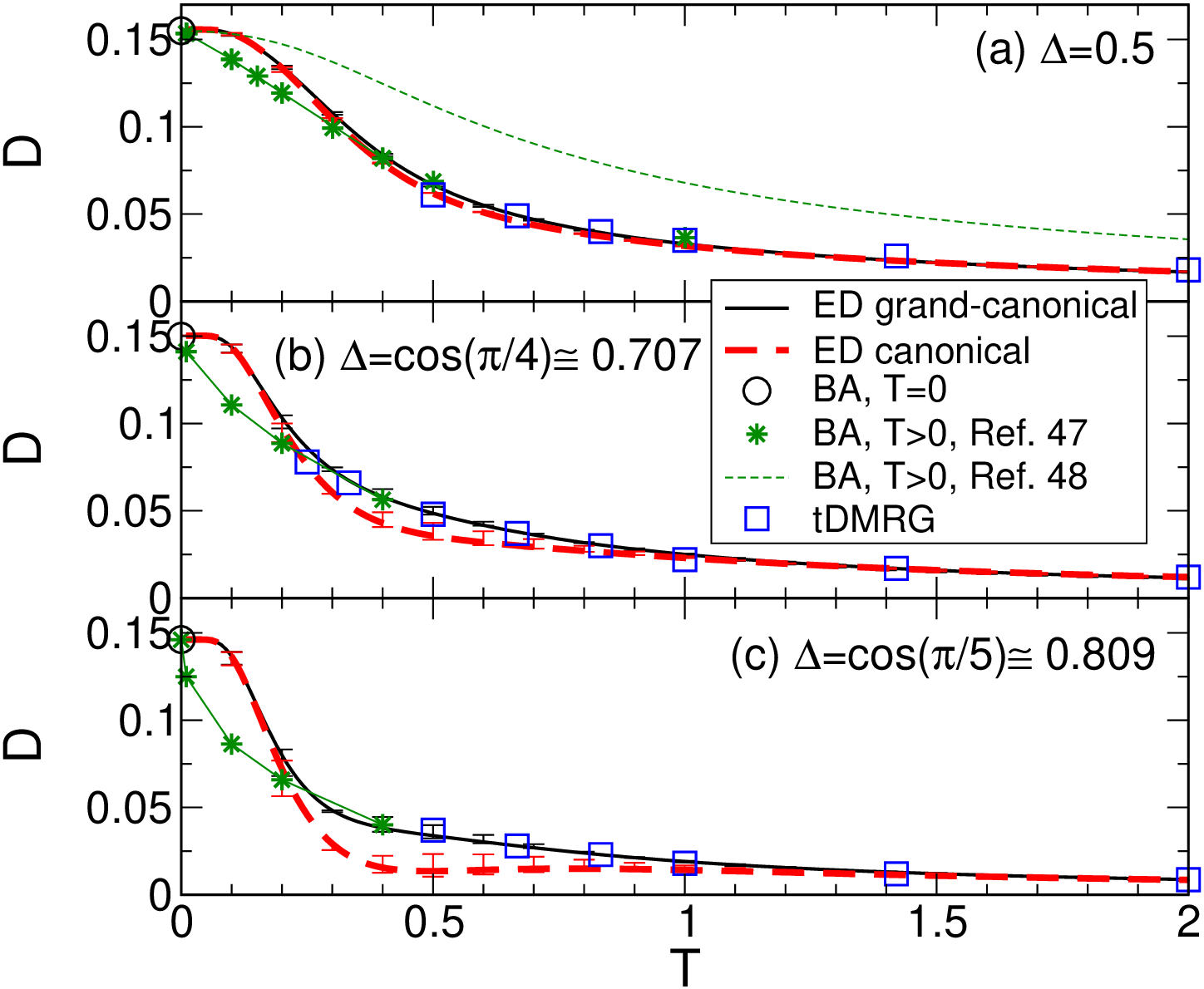}
\caption{
(Color online) Comparison of tDMRG and ED data (extrapolated in $L$) for the Drude weight for (a) $\Delta=0.5$, (b) $\Delta=\cos(\pi/4)$, and (c) $\Delta=\cos(\pi/5)$.
Circles are the exact $T=0$ Bethe ansatz results taken from Ref.~\onlinecite{shastry90}, the stars are the $T>0$ Bethe ansatz results from Ref.~\onlinecite{zotos99}, the dashed
line in (a) are the Bethe ansatz results from Ref.~\onlinecite{benz05}.}
\label{fig:D_T_posDelta}
\end{figure}

The main goal of our work is to obtain $D=D(T)$ from exact diagonalization by means of an extrapolation in $1/L$, in  order to compare with tDMRG data.
In Fig.~\ref{fig:ED_extN_D}, we demonstrate, for   the example of $\Delta=0.5$, that  the fitting function  
\begin{equation}
D(T,L) =D(T) +A(T)/L + B(T)/L^2
\label{eq:d_l} 
\end{equation}
describes the available data relatively well at {\it all} temperatures, which has previously not been appreciated in 
exact diagonalization studies (Herbrych {\it et al.} also present results for $D(T)$ from such extrapolations of canonical data as a function of $T$ in Ref.~\onlinecite{herbrych11}).
However,  in the temperature regime of $0.1<T\lesssim 0.5$, the monotony behavior of $D(T,L)$ as a function of $L$ changes
dramatically, which is particularly evident by comparing data from odd to even $L$ (see the discussion and figures in Ref.~\onlinecite{hm03}), rendering this regime the most difficult one and the least reliable for finite-size extrapolations. In Ref.~\onlinecite{herbrych11}, it was observed that $D= C/T$ approximates the full $D(T)$ well down
to temperatures of $T\sim 0.5$, suggesting that it should be possible to obtain an estimate for $D(T)$ from finite-size extrapolations in the regime $0.5<T<\infty$.
As $T\to 0$, we observe $A\to 0$ (see the discussion in Ref.~\onlinecite{hm03} and references therein).
In the following, we will use Eq.~\eqref{eq:d_l} to obtain $D(T)$ for several $\Delta=\cos(\pi/\nu)$, with $\nu=3,4,\dots,25$ and at $\Delta=1$ (corresponding to $\nu\to\infty$).  
Error bars in the figures are obtained from excluding either the largest or smallest system size from the fits. We have also tried other possible extrapolation functions, yet we 
find that a second order polynomial in $1/L$ consistently approximates the data the best.

\section{Results and comparison of tDMRG with exact diagonalization}
\label{sec:results}
In this section, we present our main results, namely ED and tDMRG data for the temperature dependence
of $D(T)$, a comparison of canonical versus grand-canonical ED data, and we revisit the case of $\Delta=1$,
for which no exact results are available on whether $D(T)$ is finite or not.
 Note that results for $D(T)$ from an extrapolation of canonical ED data were previously presented in Ref.~\onlinecite{herbrych11} (we quantitatively agree with their data for $\Delta=0.5$),
while grand-canonical ED data for finite systems was extensively discussed in Ref.~\onlinecite{hm03} (see also further references mentioned in that work).
 
\subsection{Gapless regime:  $|\Delta|<1$}
\label{sec:gapless}

Figure~\ref{fig:D_T_posDelta} contains representative examples of our results for $D(T)$ for three positive values of 
$\Delta=\cos(\pi/\nu)$ ($\nu=3,4,5$). 
We emphasize the main observations. First, the 
canonical and grand-canonical ED data agree well with each other both at low and high temperatures.
At low temperatures, the subspace with the smallest value of $S^z$ dominates the grand-canonical sum in Eq.~\eqref{eq:drude}
and since this smallest $S^z$-subspace enters in the canonical calculation of $D$, the agreement at low temperatures is 
to be expected. Second, the deviations between the canonical and grand-canonical data are the largest at intermediate temperatures
$T\sim 0.5$, and increase as $\nu$ increases. Third, the agreement between the tDMRG data and the ED data is excellent, and the 
tDMRG appears to be closer to the grand-canonical ED curves. In all cases, the extrapolated ED data are consistent with the exactly known zero-temperature result\cite{shastry90} (circles in Fig.~\ref{fig:D_T_posDelta}).

\begin{figure}[t]
\includegraphics[width=0.95\linewidth,clip]{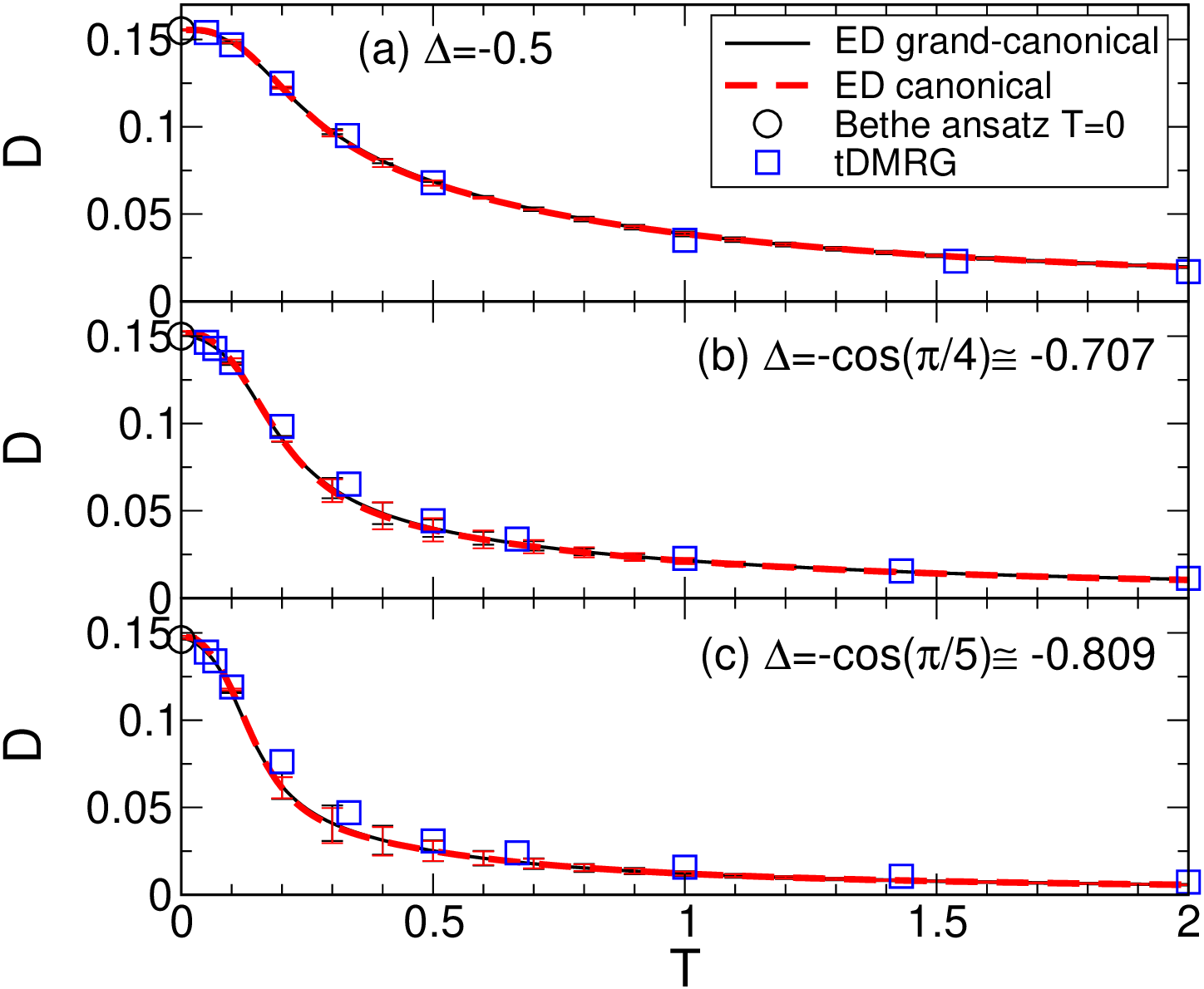}
\caption{(Color online)
Comparison between tDMRG and ED data  for (a) $\Delta=0.5$, (b) $\Delta=-0.707$ and (c) $\Delta=-0.809$.}
\label{fig:D_T_negDelta}
\end{figure}

Our grand-canonical data further agree with the available finite-temperature Bethe ansatz results from Ref.~\onlinecite{zotos99} at $T\geq 0.4$ (while the canonical data 
exhibit larger deviations). Yet, upon lowering temperature, these Bethe ansatz results are systematically below the ED data and follow
a quantitatively different temperature dependence with a non-zero slope at $T=0$ (see also Ref.~\onlinecite{fujimoto03}).

Our ED data suggest a vanishing slope of $D(T)$
as $T\to 0$ at $\Delta=0.5$, in agreement with the results of a different finite-temperature Bethe ansatz calculation.\cite{benz05}
For $T>0.1$, the results by Benz {\it et al.} \cite{benz05} [dashed line in Fig.~\ref{fig:D_T_posDelta}(a)] lie systematically above 
the results of all other methods and therefore put too much weight into the Drude weight, compared to finite-frequency
contributions (see also the discussion in Ref.~\onlinecite{sirker11}). In fact, for $\Delta=0.5$,  the data of Ref.~\onlinecite{benz05}
for $D(T)$ is approximately equal to the kinetic energy.

 The discrepancy between ED and Ref.~\onlinecite{zotos99} can at present not be resolved
yet may be amenable to alternative tDMRG methods for finite temperature.\cite{white09,barthel13a,barthel13b}
Our tDMRG results for $T=0.1$ at $\Delta=0.707$ (see Fig.~\ref{fig:trick}) 
agree better with the ED data yet the time-dependent tDMRG still exhibits a noticeable slope at the longest times reached in the simulations.
Note, though, that {\it if} the Drude weight at $\Delta=1$ is finite, then its zero-temperature slope $dD/dT$ has been predicted to be {\it positive}.\cite{benz05,fujimoto03,sirker09,hm03}

As examples for the behavior of $D(T)$ in the ferromagnetic, critical regime $-1<\Delta<0$, we present results
for $\Delta=-\cos(\pi/\nu)=-0.5,-0.707,-0.809$ ($\nu=3,4,5$) in Fig.~\ref{fig:D_T_negDelta}. In these cases, we were able to extract accurate tDMRG data down to
very low temperatures (compare the discussion of Sec.~\ref{sec:tdmrg} and the real-time data presented in Fig.~\ref{fig:params}).
The agreement between ED data on the one hand -- both canonical and grand-canonical data -- and tDMRG on the other hand is excellent, at all
temperatures considered. Small deviations occur at temperatures $T\sim 0.3$ for $\Delta=-0.809$ where the error bars of the extrapolation
of the ED data are the largest.

\begin{figure}[t]
\includegraphics[width=0.95\linewidth,clip]{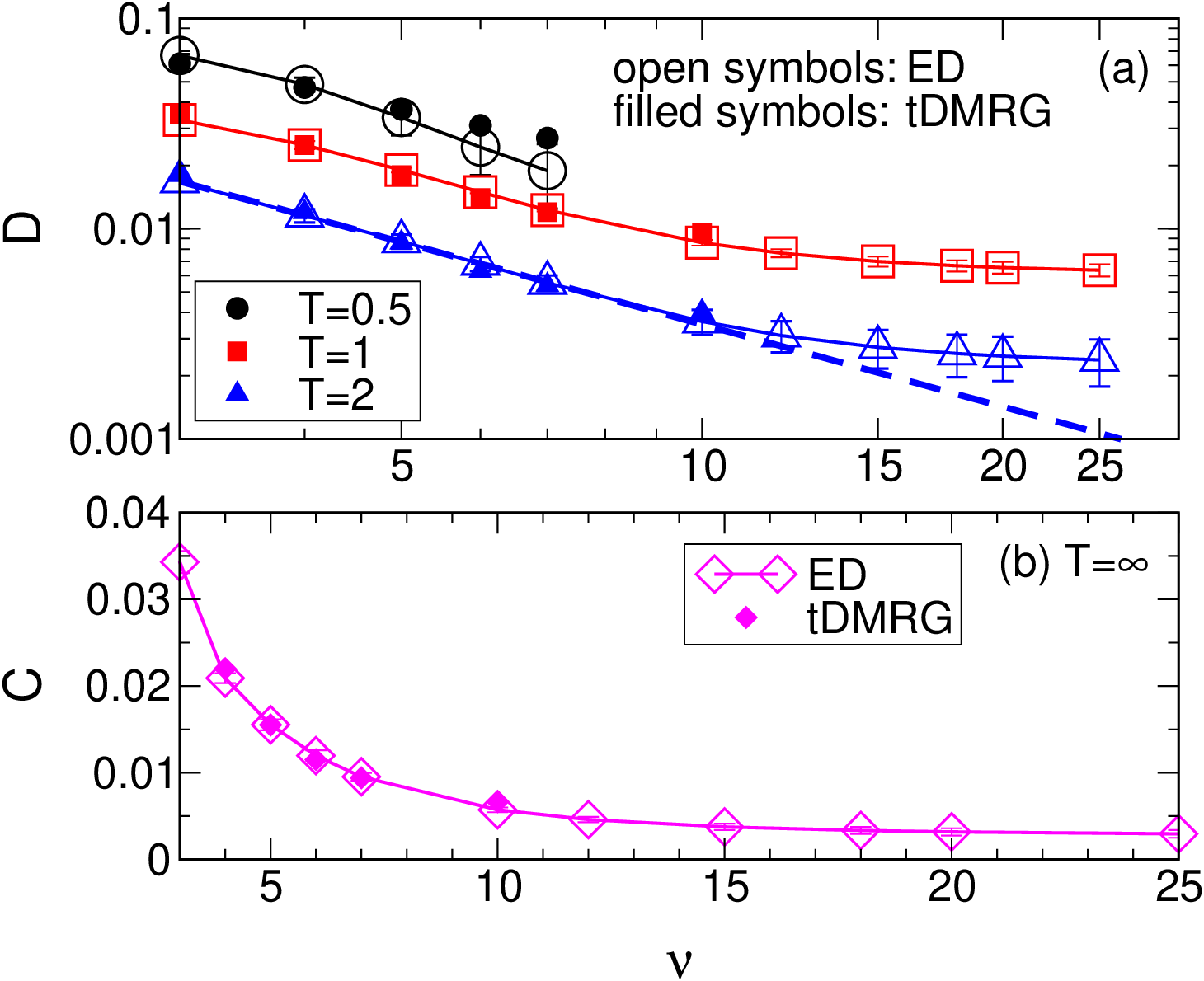}
\caption{
(Color online) (a) Drude weight versus $\nu$, where $\Delta = \cos\left(\frac{\pi}{\nu}\right)$. (a), (b) Comparison of tDMRG (filled symbols; shown for $\nu\leq10$) and ED (open symbols) using grand-canonical data
in a log-log plot. Thin solid lines connecting the ED data are guides to the eyes.
(b) $C$ vs $\nu$ ($C=\lim_{T\to \infty} \lbrack T \, D(T)\rbrack$). 
The thick dashed line in  (a) shows an extrapolation using $D(\nu) = A \cdot (\nu)^{-B}$ using $\nu = 3,\dots,7$.}
\label{fig:D_nu}
\end{figure}

We next study how $D(T)$ behaves as $\Delta=1$ is approached, sending $\nu$ to large values.
We compute $D(T,\nu)$ from grand-canonical ED data for various values of $\nu$ from an extrapolation of $D(T,\nu,L)$ in system size
that is carried out for each pair of $T,\nu$, along the lines discussed in Sec.~\ref{sec:ed} (see Fig.~\ref{fig:ED_extN_D}).
Thus, for each temperature $T$, we obtain a series of data points $D(T,\nu)$ as a function of $\nu$, where $\Delta\to 1$ as
$\nu\to \infty$. Examples are shown in Fig.~\ref{fig:D_nu}(a). For $\nu\leq 10$, we include tDMRG data  
that is in reasonable qualitative agreement with the ED results (for $\nu > 10$,  we cannot reliably extract results for $D(T)$ from the real-time tDMRG simulations). 
The figure further shows that the ED data do not follow a power-law in $\nu$ (dashed line in the figure), but rather, $D(T,\nu)$ seems to level off at finite and infinite temperature, suggesting a non-zero value
of $D(T)$ at $\Delta=1$. Based on the available real-time DMRG data for $\Delta=\pm 1$ (Figs.~\ref{fig:corr1} and \ref{fig:d1m}), we cannot unambiguously discern between a finite Drude weight at $\Delta=1$ and a current correlation function that decays on a fairly large time scale.

\subsection{Spectral weight of the Drude peak}
\label{sec:weight}
Finally, we study how much of the total weight $I_0$ of the spin conductivity is in the Drude weight, compared to finite-frequency contributions.
A similar analysis of ED data  was presented in Ref.~\onlinecite{herbrych11} using canonical ED data at $T=\infty$ exploring the dependence on $\Delta$ and $S^z$, which we here complement by 
using grand-canonical data and consideration of finite temperatures $0<T<2$.

\begin{figure}[t]
\includegraphics[width=0.95\linewidth,clip]{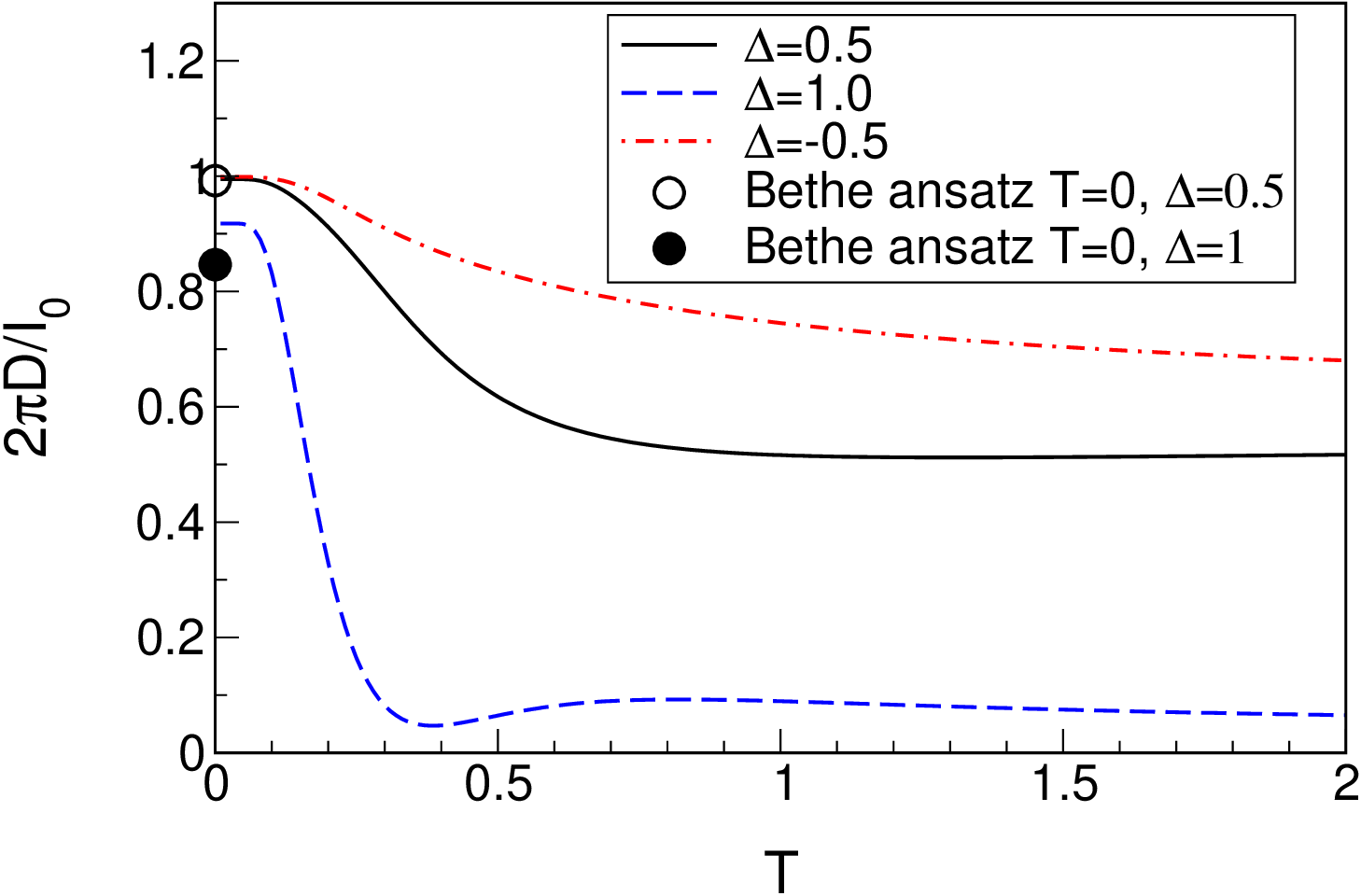}
\caption{
(Color online)
Fraction of the Drude weight $2 \pi D /I_0$ to the total spectral weight of the spin conductivity, $I_0$, obtained from ED by grand-canonical calculation of Eq.~\eqref{eq:drude}.
 $I_0 = \int \tn{Re } \sigma (\omega) d\omega$, through the f-sum rule given by\cite{shastry90} $I_0 = \frac{\pi}{L} \langle-\hat{T}\rangle$.
The extrapolation to $L\rightarrow\infty$ is done as in Fig.~\ref{fig:ED_extN_D}.   }
\label{fig:ED_D_over_I}
\end{figure}

The real part of the spin conductivity obeys a sum rule:
\begin{equation}
I_0=\int d\omega\, \mbox{Re}\,\sigma(\omega)= \frac{\pi}{L} \langle -\hat T\rangle\,. 
\end{equation}
We plot $2\pi D/I_0$ in Fig.~\ref{fig:ED_D_over_I} as a function of temperature for $\Delta=-0.5,0.5$ and $1$.
 These data are obtained from grand-canonical ED results plus an extrapolation
using Eq.~\eqref{eq:d_l}. The units are chosen such that the vertical axis can be read in \%, implying that $2\pi D/I_0=1$ corresponds to the absence 
of any finite-frequency contributions.

At $\Delta=\pm 0.5$ and in the low temperature limit, practically all the weight sits in the Drude weight which then monotonously
drops to values larger than 50\% of the total weight at high temperatures (at $T=\infty$, the values for $\Delta=0.5$ and $-0.5$ should agree with each other).\footnote{$C(\Delta)=C(-\Delta)$ is strictly correct for even $L$ and in the thermodynamic limit, yet not for finite, odd $L$. Since our extrapolations include small odd $L$, we recover 
$C(\Delta)=C(-\Delta)$ only approximately as $L\to \infty$.}  
By all means, the contribution of the Drude weight is therefore substantial
at all temperatures. 
At $\Delta=1$, the low-temperature regime is difficult to obtain from ED data due to several reasons.
It is well-known that the low-temperature behavior of thermodynamic quantities at $\Delta=1$ is strongly
influenced by the appearance of a marginally irrelevant perturbation to the effective low-energy field theory, which otherwise, in the
gapless phase, is simply a Luttinger liquid.  This cannot be captured by ED on short chains (compare the work by Laflorencie {\it et al.}, Ref.~\onlinecite{laflorencie04}),
and we thus expect the temperature dependence of the true Drude weight to deviate from our result at low temperatures $T\lesssim 0.1$.
 The exact $T=0$ result
for $2\pi D/I_0$ (circle) sits well below the ED data.\cite{hm03} Based on the available ED data, we conjecture that one possible 
behavior of $D(T)$ and $2\pi D/I_0$ in the thermodynamic limit at $\Delta=1$ is a {\it maximum} at some intermediate $T\sim 0.1$ if $D(T)$ is non-zero at all.
This would imply a positive slope of $D(T)$ at zero temperature, in agreement with Refs.~\onlinecite{fujimoto03,benz05}.
The existence of the additional maximum in $D(T)$ at $T\sim 0.8$ cannot be resolved with the available system sizes and should thus
be considered as spurious.
 An intriguing observation from Fig.~\ref{fig:ED_D_over_I} for $\Delta=1$ is that
the Drude weight amounts to 80\% of the total weight at temperatures $T\sim 0.1$, which drops to less than 10\% at high temperatures.

Our results indicate that the contribution of the Drude peak is
substantial for all $\Delta>0$ and $\Delta$ sufficiently smaller than $1$,
and may even be large at low temperatures (see the Fig.~\ref{fig:ED_D_over_I} with $\Delta=0.5$).
We further suggest that our data for the temperature dependence of $2\pi D(T)/I_0$ at $\Delta=\pm0.5$ could be used to
obtain estimates on the diffusion rate $\gamma$ introduced in Refs.~\onlinecite{sirker09,sirker11} along the lines of Ref.~\onlinecite{sirker11} or for a comparison with QMC results.\cite{grossjohann10} 
Finally, we observe that for the commensurate values of $\Delta=\cos(\pi/\nu)$ studied in our work, the lower bound to $D(T)$ from Prosen's work\cite{prosen11} seems to
be almost exhaustive.

\section{Summary}
\label{sec:sum}

In this paper we studied the problem of the spin Drude weight of the spin-$1/2$ $XXZ$  chain at finite temperature using the time-dependent density matrix renormalization group as well as exact diagonalization. This complements earlier works in various ways. First, we elaborated on numerical details of both methods. tDMRG yields the real-time current-current correlation function $\langle J(t)J(0)\rangle$, and $D$ can be extracted from its asymptote provided that the asymptotic time regime where $\langle J(t)J(0)\rangle$ saturates can be reached. We presented data for various discarded weights which controls the accuracy of tDMRG calculations. While the thermodynamic limit $L\to\infty$ can be accessed easily within tDMRG, exact diagonalization is bound to system sizes $L\lesssim20$ and thus requires an extrapolation to $L\to\infty$. We showed that canonical and grand-canonical ED
data are in good agreement at small $\Delta <1$. Grand-canonical data appear to have more systematic finite-size dependencies all across the gapless phase.  As a second key aspect of our work, we presented extensive data for the temperature-dependence of the Drude weight. For parameters where the asymptote of $\langle J(t)J(0)\rangle$ can be accessed by tDMRG (antiferromagnetic chains at intermediate-to-high temperatures as well as for ferromagnetic chains at any $T$), results are in perfect agreement with ED extrapolated to $L\to\infty$. 
Finally, we investigated how $D(T)$ behaves in the vicinity of $\Delta=1$
by sending $\nu$ to large values ($\Delta=\cos(\pi/\nu)$). These data are consistent with $D(T)>0$ at $\Delta=
1$ (in disagreement with Refs.~\onlinecite{zotos99,herbrych11}), and a final clarification of this open question is left for  future work, exploiting, for instance, 
  alternative and recently developed finite-temperature tDMRG methods.\cite{white09,barthel13a,barthel13b}

\emph{Acknowledgments} --- We thank W. Brenig, J. E. Moore, T. Prosen, and F. Verstraete for very helpful discussions and we thank A. Kl\"umper for his comments on a previous version of the manuscript and for sending us data from Ref.\onlinecite{benz05}. We gratefully  acknowledge  support from to the Deutsche Forschungsgemeinschaft through  grant-no. 
KA3360-1/1 (C.K.) and through  Research unit FOR 912 (grant no. HE-5242/2-2 (J.H. and F.H.-M.)) as well as from the Nanostructured Thermoelectrics program of LBNL (C.K.).


\bibliographystyle{apsrev}
\bibliography{references}

\end{document}